\documentclass[prl,aps,twocolumn,showpacs]{revtex4-1}
\include{graphics}
\usepackage{epsfig}
\usepackage{graphicx}
\usepackage{textcomp}
\begin{document}
\title{The dependence of the dry friction threshold on rupture dynamics}
\author{Oded Ben-David and Jay Fineberg}
\affiliation{The Racah Institute of Physics, The Hebrew University
of Jerusalem, Jerusalem 91904, Israel}
\begin{abstract}
The static friction coefficient between two materials is considered to be a material constant. We present experiments demonstrating that the ratio of shear to normal force needed to move contacting blocks can, instead, vary systematically with controllable changes in the external loading configuration. Large variations in both the friction coefficient and consequent stress drop are tightly linked to changes in the rupture dynamics of the rough interface separating the two blocks.
\end{abstract}
 \maketitle

For centuries the onset of frictional motion has been described by the concept of a friction coefficient, $\mu$, that reflects proportionality between applied shear and normal forces when frictional motion initiates.  This concept, which dates back to DaVinci, Amontons, and Coulomb, was first explained by Bowden and Tabor \cite{bowden_friction_2001} as resulting from the proportionality of the real area of contact, $A$ with the normal force, $F_N$, pressing two bodies together. Phenomenological corrections describing the dependence of the friction coefficient on time and sliding velocity were later introduced  \cite{ruina_slip_1983,*dieterich_time-dependent_1978,*marone_laboratory-derived_1998,*scholz_earthquakes_1998,*persson_nature_2003,berthoud_physical_1999,ben-david_slip-stick_2010,rice_rate_2001}.
Although significant, these corrections generally predict variations of $\mu$ of a few percent.

The onset of frictional slip is, however, caused by the rapid rupture of the discrete ensemble of contacts that forms a rough frictional interface \cite{rubinstein_detachment_2004,ohnaka_characteristic_1990,ben-david_dynamics_2010,ben-david_slip-stick_2010,xia_laboratory_2004,*andrews_rupture_1976,*burridge_admissible_1973,*das_spontaneous_2003,rice_rate_2001}. The nature of these rupture modes is coupled to the local profiles of the shear, $\tau(x)$, and normal, $\sigma(x)$, stresses along the interface with local variations in $\tau(x)/\sigma(x)$ significantly influencing the rupture mode selection \cite{ben-david_dynamics_2010}. Moreover, for given external loading conditions $\tau(x)/\sigma(x)$ can not only vary considerably but can, locally, surpass macroscopic friction coefficient, $\mu$, by hundreds of percent \cite{ben-david_dynamics_2010}, demonstrating that $\mu$ is not the local criterion for slip onset.

We will demonstrate that the static friction coefficient, $\mu_S$, is {\em not} a constant material property. It, in fact, varies systematically by nearly a factor of two with the external loading configuration. Furthermore, the threshold for frictional sliding is intimately linked to the rupture dynamics of the interfacial contacts. This link provides a mechanical explanation for significant variations of  $\mu_S$ found even within a single experiment with a set external loading configuration.

\begin{figure}[ht]
\includegraphics[width=0.95\columnwidth,clip=true,keepaspectratio=true]{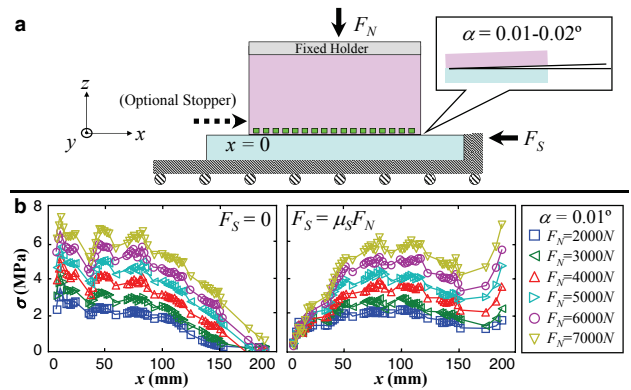}
\caption{(a) Schematic view of the experimental setup. Normal force $F_N$ is applied uniformly to the top block. Shear force $F_S$ is applied to the translational stage parallel to the $x$ axis. Stresses were measured along the frictional interface by miniature strain-gages (green squares, not to scale).  A small tilt, $\alpha$, was introduced to controllably vary the stress profiles (call-out). (b-left) The initial $\sigma(x)$ profiles for $\alpha=0.01^\circ$ and $F_N=2000-7000N$ (see legend), showing how stress profiles differ with $F_N$. (right) $\sigma(x)$ profiles immediately prior to slip initiation for events from experiments with these values of $F_N$ (color coding as in (b)). For each $F_N$, the initial $\sigma(x)$ profiles are changed to a different degree by the $F_S$ induced torque.}\label{fig1}
\end{figure}

 The experiments described here were all in the stick-slip regime, where each experiment consisted of a series of stick-slip events. The experimental system, depicted schematically in Fig. \ref{fig1}, is similar to that described in \cite{ben-david_slip-stick_2010,ben-david_dynamics_2010}. Two PMMA blocks (shear wave speed, $C_S=1370m/s$) were pressed together with a uniform normal force $F_N$. The top (bottom) block  dimensions were $200\times 6\times 100mm$ ($300\times 60\times 30mm$), where  $x$, $y$ and $z$ are, respectively, the propagation, thickness and normal loading directions. The bottom block was mounted on a low-friction translational stage constrained to move along the $x$ axis. Relative changes in the real contact area along the rough interface, $A(x,t)$, during each stick-slip event were measured by a method based on total internal reflection \cite{rubinstein_detachment_2004,ben-david_slip-stick_2010,ben-david_dynamics_2010}, where $t$ denotes the time following the rapid rupture initiation for each event. $A(x,t)$ are normalized by values $1ms$ prior to rupture initiation at $t=0$.  
 
Shear force, $F_S$, was applied to the bottom block in the negative $x$ direction via a load-cell of stiffness $10^7N/m$ at a loading rate of $10-20N/s$. $F_N$ and $F_S$ were continuously measured at $100Hz$. In addition, $F_S$ was measured at $250KHz$ during stick-slip events to accurately determine its peak value for each event (each event corresponded to a sharp drop in $F_S$). Force ratios $\mu(t)=F_S(t)/F_N(t)$, were measured to better than $1\%$ accuracy. For each event, we define $\mu_S\equiv\mu(t=0)$. Rapid acquisition of $A(x,t)$ and $F_S$ was triggered by an acoustic sensor mounted to the $x=0$ face of the top block. Local strain values were measured $\sim 2mm$ from the interface \cite{ben-david_dynamics_2010}, yielding profiles of $\sigma(x)$ and $\tau(x)$ prior to each slip event, at 56 and 8 locations along the interface, respectively.

To controllably vary both the stress profiles along the interface and the resulting rupture dynamics, we varied the loading conditions by slightly tilting the top block at angles $\alpha=0.01^\circ$ and $\alpha=0.02^\circ$ (see Fig. \ref{fig1}(a)). Note that while the absolute value of $\alpha$ is accurate to within $10\%$, its value was reproducible to within $1\%$ in sequential experiments. For a full discussion of loading see \cite{ben-david_dynamics_2010}. For a given $\alpha$, variation of  $F_N$ over the range $2000-7000N$ controllably changes the stress profiles. This is illustrated in Fig. \ref{fig1}(b-left) which shows how $\sigma(x)$ changes as $F_N$ is increased for $\alpha=0.01^\circ$. With $\alpha$ in this orientation, these changes in $\sigma(x)$ are partially offset by the torque applied by $F_S$. Fig. \ref{fig1}(b-right) shows how $F_S$ influences $\sigma(x)$ profiles at $t=0$. $\sigma(x)$ profiles vary significantly for different $F_N$ at $t=0$ due to different degrees of compensation by the $F_S$-induced torque. Increasing $\alpha$ changes both the initial profiles and resulting stress profiles at slip initiation.

\begin{figure}[ht]
\includegraphics[width=0.95\columnwidth,clip=true,keepaspectratio=true]{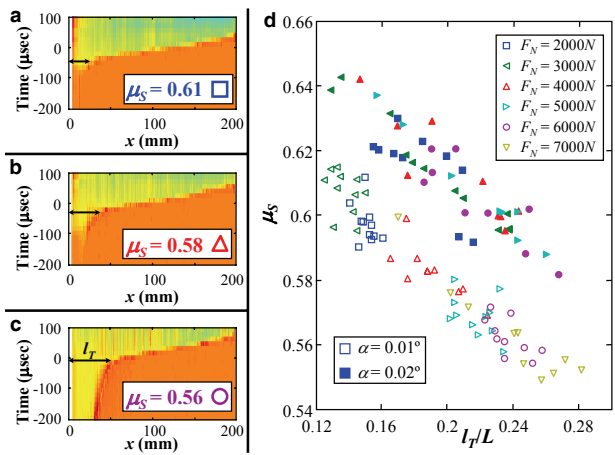}
\caption{(a-c) Measurements at $20{\mu}s$ intervals of the changes in the real contact-area $A(x,t)$ for representative events with $\alpha=0.01^\circ$ and (top-bottom) $F_N=2000, 4000$ and $6000N$. Hotter (colder) colors indicate increased (decreased) $A$. $A(x,t)$ was normalized by its value $1ms$ before rupture initiation. Each rupture consists of a slow front initiating near $x=0$ which sharply transitions, at $x\equiv l_T$ to rapid propagation. Transitions to front velocities $V = 2000, 1800$ and $1500 m/s$ occur at $l_T = 27, 38$ and $52mm$ for (a-c). $\mu_S$ decreases with increasing $l_T$. (d) $\mu_S$ vs. $l_T$ normalized by the interface length $L=200mm$, for 122 events with $\alpha=0.01^\circ$ (open symbols) and $\alpha=0.02^\circ$ (filled symbols) for $F_N=2000-7000N$. For each $\alpha$, all points collapse to distinct well-defined curves, each offset in $\mu_S$. Whereas data for $\alpha=0.01^\circ$, values of $\mu_S$ roughly cluster for each $F_N$, data for both $F_N=7000N$ and $\alpha=0.02^\circ$ span most of the collapse curves, indicating that $\mu_S$ depends on rupture dynamics rather than directly on $F_N$.
} \label{fig2}
\end{figure}

In accepted views of static friction, $\mu_S$ is a material property which should {\em not} depend on $F_N$.
Fig. \ref{fig2} demonstrates that $\mu_S$, in fact, changes significantly with the loading details.
Furthermore, the variation of $\mu_S$ is systematically correlated with the dynamics of the interface rupture.
Examples of how rupture dynamics are affected by changing $F_N$ for initial values of $\alpha$, are presented in Fig. \ref{fig2}.
The left panel presents $A(x,t)$ measurements for three representative slip events.
The front dynamics for all events are qualitatively similar; a slow front ($30 \leq V \leq 300 m/s$) initiates near $x=0$ and undergoes a sharp transition to rapid rupture ($1300 \leq V \leq 2000m/s$) which traverses the rest of the interface. These dynamics are consistent with the ``nucleation phase" of \cite{ohnaka_characteristic_1990}.
Quantitatively, however, the length of each slow front at transition, denoted by $l_T$, varies with the changing stress profiles for different $F_N$. The resultant $\mu_S$ systematically decrease as $l_T$ grows (Fig. \ref{fig2}(a-c)).
In Fig. \ref{fig2}(d) $\mu_S$ versus $l_T$ (normalized by the interface length, $L$) is presented, for both $\alpha=0.01^\circ$ and $\alpha=0.02^\circ$, over a range of $F_N$. Included are 122 events from 11 experiments.

The data for $\alpha=0.01^\circ$ roughly cluster around a different value of $\mu_S$ for each value of $F_N$.  $\mu_S$, however, does {\em not} explicitly depend on $F_N$. A close look at $\mu_S$ for both $F_N=7000N$ and $\alpha=0.02^\circ$, reveals that these data do not cluster for specific values of $F_N$, but instead span nearly the entire range of $\mu_S$. When plotted as a function of $l_T$, the ``scattered" data from {\em all} experiments collapse onto well-defined curves with very little scatter for each $\alpha$. While the two data sets are offset in $\mu_S$, events with the same $\alpha$ have similar slopes. Thus, the systematic variation of $\mu_S$ is not an explicit dependence on $F_N$, but is, instead, closely tied to the rupture dynamics.

For these loading conditions, the rupture dynamics of the first event in each experiment generally differ from those of subsequent events.
Stresses along the interface are induced by Poisson expansion that is frustrated due to frictional pinning \cite{rubinstein_contact_2006,ben-david_slip-stick_2010}. These stresses often lead, in this configuration, to rupture nucleation of the first event away from the sample edges (c.f. Fig. \ref{fig4}).
These data were, therefore, not included in Fig. \ref{fig2} since, in these cases, $l_T$ is not defined.

\begin{figure}[ht]
\includegraphics[width=0.95\columnwidth,clip=true,keepaspectratio=true]{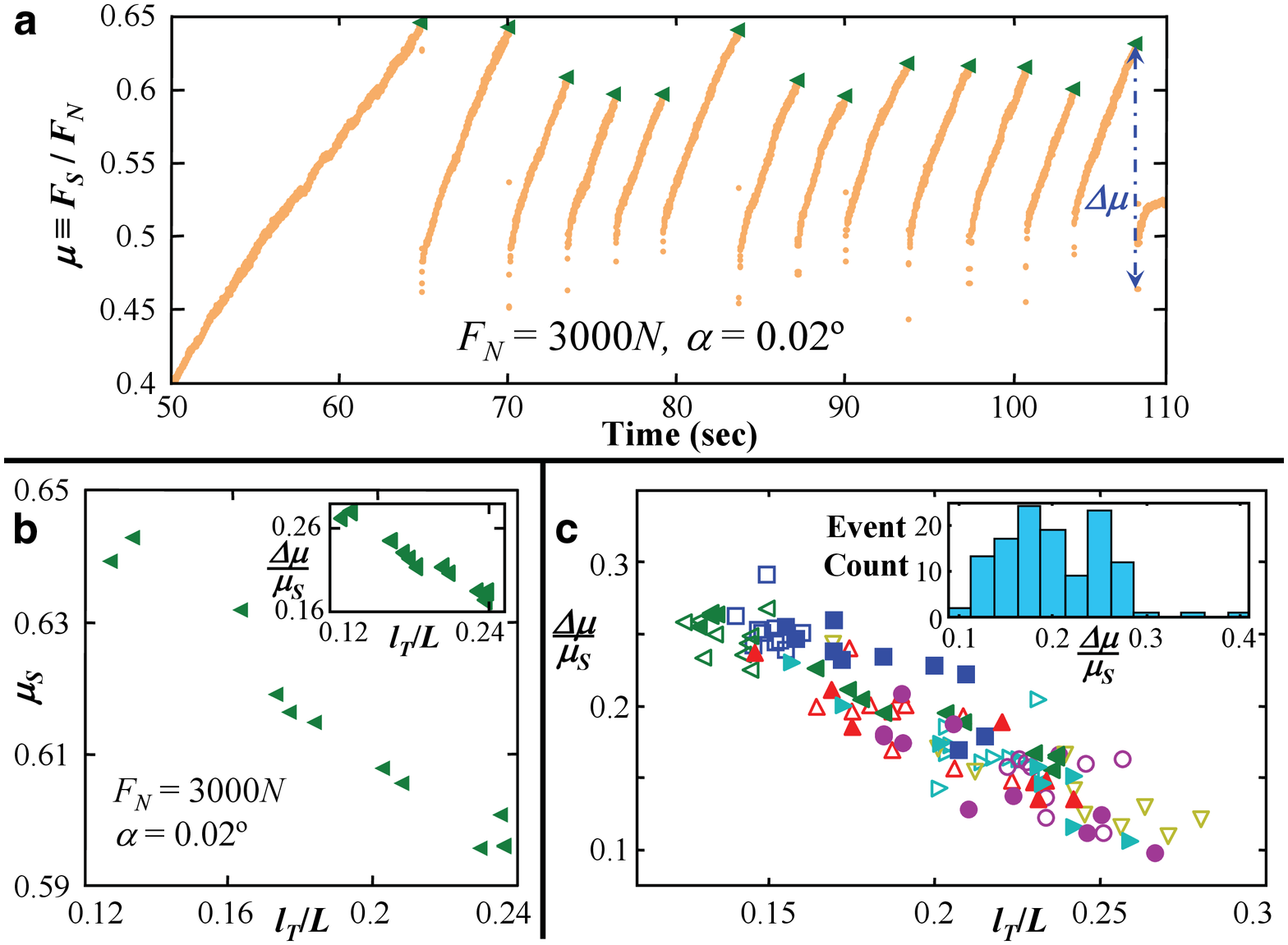}
\caption{(a) The loading curve $\mu$ as a function of time for $F_N=3000N$ and $\alpha=0.02^\circ$. $\mu(t)$ drops by $\Delta \mu$ after each event. Values of $\mu_S$ (triangles) vary by up to $8\%$ in subsequent events. $\mu_S$ and the relative stress drops, $\Delta \mu /\mu_S$, collapse to well-defined curves when plotted vs. $l_T/L$. (b) $\mu_S$ vs. $l_T/L$ (inset) $\Delta \mu /\mu_S$ vs. $l_T/L$ for the data in (a). (c) $\Delta \mu /\mu_S$ vs. $l_T/L$ for the broadly distributed $\Delta \mu /\mu_S$ (inset). Plotted are all data from $\alpha=0.01^\circ- 0.02^\circ$ presented in Fig. \ref{fig2}d.} \label{fig3}
\end{figure}

\begin{figure}[ht]
\includegraphics[width=0.95\columnwidth,clip=true,keepaspectratio=true]{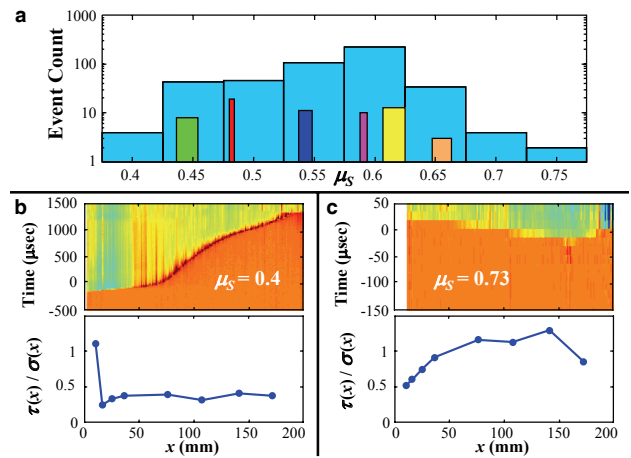}
\caption{(a) Count of events with varying $\mu_S$, from experiments conducted under widely varying loading conditions. These range from pure edge loading to near uniform application of $F_S$. Note that this plot is {\em not} a histogram and solely reflects the number of experiments performed whose loading conditions yielded $\mu_S$ for each range. The color bars (bar width = two standard deviations) in the plot depict sub-populations of events from single experiments, each with different loading conditions. The bar at $\mu_S=0.62$ corresponds to Fig. 3(a-b). (b-c) $A(x,t)$ (top) and stress ratio profile immediately prior to the event (bottom) for two experiments at extreme values of $\mu_S$. (b) Applying $F_S$ to the edge of the sample resulted in strong localization of $\tau$ near the edge, and initiation of a slow ($\sim 90m/s$ on average) rupture which traversed most of the sample, resulting in $\mu_S=0.4$. (c) The relatively uniform stress-ratio profile with lower stress-ratio near the edges inhibited rupture initiation, resulting in a supershear rupture initiating at $x \approx 158mm$ with $\mu_S=0.73$.} \label{fig4}
\end{figure}

Large variations in both the peak values of $\mu$ ($\mu_S$) and the consequent drops, $\Delta \mu$, in each stick-slip event are apparent in the representative time series of successive slip events presented in Fig.\ref{fig3}(a). $\mu_S$ variations are notably larger ($\lesssim 8\%$) for $\alpha=0.02^\circ$ than for $\alpha=0.01^\circ$ ($\lesssim 3\%$) for the same $F_N$. The size of these variations are not explained, for example, by logarithmic aging in PMMA, $\mu_S(t)=\mu_S(0) + 0.01log(t)$ \cite{berthoud_physical_1999,ben-david_slip-stick_2010}, which accounts for only a $\sim 0.4\%$ variation in $\mu_S$ over the temporal-interval variations between slip events. In general, distributions of both $\mu_S$ and $\Delta \mu /\mu_S$ are quite broad, as demonstrated in Fig.\ref{fig3}. This apparent random aspect of frictional stability disappears when both $\mu_S$ and $\Delta \mu /\mu_S$ are plotted as a function of $l_T$. In particular, Fig. \ref{fig3}(c) demonstrates that the broad distribution of $\Delta \mu /\mu_S$ (Fig. \ref{fig3}(c-inset)), obtained for all experiments with both $\alpha$ values, collapses onto a relatively well-defined curve, when plotted as a function of $l_T$. Thus, the aperiodic nature of stick-slip friction \cite{johnson_effects_2008,*capozza_rubinstein_2011} is also closely related to rupture dynamics.

To this point we have seen how controlled changes in the loading configuration (either different $F_N$ and/or $\alpha$) can, by varying the rupture dynamics, induce systematic changes in $\mu_S$ of order 10\%.
To demonstrate the dependence of $\mu_S$ on rupture dynamics, we used particular examples of loading configurations in which rupture dynamics could be roughly characterized by a single parameter (e.g. $\l_T$).

Even in these simple cases, $\l_T$ is not uniquely determined by details of the loading (c.f. Fig. 2). In the general case, moreover, the stress profiles and ensuing rupture dynamics are strongly dependent on the loading configuration \cite{ben-david_dynamics_2010} and are not, necessarily, characterized by a single parameter. In Fig. \ref{fig4}, we show that the spread in $\mu_S$ can, in fact, be much larger when a larger variation of stress profiles along the interface is imposed.
Fig. \ref{fig4} presents $\mu_S$ data for $458$ slip events that resulted from a variety of different loading configurations.
These range from pure edge loading (where $F_S$ is applied directly to a sample edge) to the loading configurations described in Figs. \ref{fig1}-\ref{fig3}.
Further details of the various configurations can be found in \cite{ben-david_dynamics_2010,ben-david_slip-stick_2010}. The figure shows that, simply by changing the loading configuration, $\mu_S$ can vary by a factor of almost two {\em for the same two blocks}.
Note that Fig. \ref{fig4}(a) is {\em not} a histogram, but simply a record of the number of events in which a given $\mu_S$ was measured.
Sub-populations of events from single experiments (colored bars in Fig. \ref{fig4}(a)) demonstrate that for given loading configurations, $\mu_S$ is relatively well-defined. For given loading conditions, the maximal spread of $\mu_S$ is close to the $\sim 8\%$ variation presented in Fig. \ref{fig3}.

In Fig. \ref{fig4}(b),(c) we present two sample slip events which correspond to the low and high extremes of $\mu_S$ within Fig. \ref{fig4}(a).
The $\mu_S=0.4$ event (Fig. \ref{fig4}(b)) was obtained by application of $F_S$ solely to the edge of the top block. This lead to a $\tau/\sigma$ profile whose maximal value was highly localized near $x=0$, and much lower elsewhere. In \cite{ben-david_dynamics_2010} it was shown that once a rupture front nucleates, ruptures will propagate in any of three different modes, as long as $\tau(x)/\sigma(x)$ locally exceeds $\sim 0.25$.
Here, a rapid front ($V \approx 0.4C_S$) nucleated at the high $\tau/\sigma$ ratio at $x=0$ and then continued at $x \approx 60mm$ as a slow rupture front ($V \sim 0.07C_S$)  across most of the interface, before transitioning back to rapid propagation as the rupture approached the $x=200mm$ edge. As most of the interface ruptured near the minimal values of $\tau/\sigma$ ($\sim 0.25$) that support rupture propagation, the overall value of $\mu_S$, which is the ratio of the {\em spatial integrals} of $\tau$ and $\sigma$, was extremely low.

In the $\mu_S=0.73$ event (Fig. \ref{fig4}(c)) the values of $\tau(x)/\sigma(x)$ across nearly the entire interface were both significantly higher ($\sim 1$) than in the $\mu_S=0.4$ event and uniform, except for the relatively low values near the edges. In this case, the rupture nucleated within the interior of the interface, at a point ($x=158mm$) where  $\tau/\sigma$ was near maximal. Once nucleated, the resulting counter-propagating ruptures sharply transitioned to super-shear propagation ($V \sim 1.85C_S$), as predicted in \cite{ben-david_dynamics_2010}. Here, the high overall value of $\mu_S$ reflects the large mean values of the $\tau/\sigma$ spatial profile. A cardinal difference between the two events in Figs. \ref{fig4}(b),(c) is in their rupture nucleation threshold.

We have demonstrated that global quantities (e.g.  $\mu_S$ and stress drops) are both strongly dependent on the imposed loading configurations and tightly linked to rupture dynamics. We suggest that the nucleation threshold for a given loading configuration may be analogous to the onset of fracture. Initiation of rapid fracture (the Griffith criterion) will occur when the amount of bulk potential energy released by extension of an existing crack surpasses the fracture energy, the energy required to create the new surfaces needed to extend the crack \cite{freund_dynamic_1990}. Fracture initiation, in any given loading configuration, is solely determined by this {\em energy} balance and {\em not} by the force balance inherent in the concept of a ``friction coefficient". In many loading configurations, the fracture threshold strongly decreases with the increased length of an initially imposed ``seed"  crack. The systematic decrease in $\mu_S$ with $l_T$ in Figs. \ref{fig2}-\ref{fig3} could suggest that the length of the slow front phase ($l_T$) plays an analogous role to this initial seed crack in determining the stability of a frictional interface to rapid rupture. The role of $\tau/\sigma$ in this process is also understood in this analogy. $\sigma(x)\propto A(x)$ is related to the local frictional strength, analogous to the fracture energy. In contrast to fracture, the frictional strength is not a material constant; it varies spatially and temporally with $A(x,t)$. $\tau(x)$ reflects the effective amount of stored potential energy surrounding a rupture tip.

The above analogy provides a conceptual framework for understanding how and why the onset of frictional slip can vary. It, furthermore, relates $\mu_S$ and rupture dynamics. It is, however, far from the entire story. Unlike fracture, where dissipative processes only occur at a crack's tip and the crack faces are stress-free, in frictional slip the ``faces" behind the rupture tip are far from free. Although the frictional strength in the wake of the rupture tip is reduced, nearly the same level (80\%) of contact area persists behind the rupture tip \cite{rubinstein_detachment_2004,ben-david_slip-stick_2010,dieterich_direct_1994} during the rupture process. A theory that fully accounts for these important effects does not yet exist.

This intuitive framework also provides a ``mechanical" explanation for widely different measured values of $\mu_S$ obtained for the {\em same} materials \cite{popov_influence_2009,berthoud_physical_1999,rubinstein_detachment_2004,ben-david_slip-stick_2010} when different loading conditions are applied. This additionally explains differences in the local and global friction coefficients observed in simplified models \cite{maegawa_precursors_2010,*scheibert_role_2010}. Dynamical changes to the stress profiles along an interface in the course of a single experiment also may explain the origin of aperiodic stick-slip behavior \cite{johnson_effects_2008,*capozza_rubinstein_2011}.

We are not yet able to predict the onset of slip for a given configuration. Our results, nonetheless, indicate that the answer may well lie in obtaining a better understanding of the evolution of the stress profiles driving the frictional rupture. This is a fundamental open question; of paramount importance to both mechanical stability and earthquake prediction.

We acknowledge the support of the US-Israel Binational fund (grant no. 2006288), the James S. McDonnell Fund and the European Research Council (grant no. 267256). We thank I. Svetlizky, G. Cohen, A. Sagy and S. M. Rubinstein for fruitful comments.
\end{document}